**Letter**

**Short-term effects of Gamma Ray Bursts on oceanic photosynthesis**

Liuba Peñate[1], Osmel Martín[2], Rolando Cárdenas[3] and Susana Agustí[4]

[1] *Department of Biology, Universidad Central de Las Villas, Santa Clara, Cuba.*
*Phone 53 42 281692 Fax 53 42 281109 e-mail: liubapa@uclv.edu.cu*

[2, 3] *Department of Physics, Universidad Central de Las Villas, Santa Clara, Cuba.*
*Phone 53 42 281109 Fax 53 42 281109 e-mail:* [2] *osmel@uclv.edu.cu;* [3] *rcardenas@uclv.edu.cu*

[4] *Department of Global Change Research, Instituto Mediterráneo de Estudios Avanzados, CSIC - UIB, Spain. Phone 34 971 611724 Fax 34 971 611761 e-mail: sagusti@imedea.uib-csic.es*

**Abstract:** We continue our previous work on the potential short-term influence of a gamma ray bursts on Earth's biosphere, focusing on the only important short-term effect on life: the ultraviolet flash which occurs as a result of the retransmission of the $\gamma$ radiation through the atmosphere. Thus, in this work we calculate the ultraviolet irradiances penetrating the first hundred meters of the water column, for Jerlov's ocean water types I, II and III. Then we estimate the UV flash potential for photosynthesis inhibition, showing that it can be important in a considerable part of the water column with light enough for photosynthesis to be done, the so called photic zone.

**1 Introduction**

Gamma Ray Bursts (GRB's) are the most luminous electromagnetic events so far discovered in the Universe. There are short and long ones, the latter being the majority, and also much better studied. The origin of long GRB's is still a debated issue, but there is compelling evidence linking them to the death of massive stars, such as core-collapse supernovae. The vast amount of gamma energy released is of around $10^{44}$ J, and it is thought to be narrowly beamed.

The estimated rate of long GRB's in the Milky Way is of about one burst every 100,000 to 1,000,000 years. Only a few percent of these would be beamed towards Earth. On the other hand, there are indications that long GRB's preferentially occur in regions of low metalicity, and because the Milky Way has been metal-rich since before the Earth formed, this effect may diminish the possibility that our planet had been stroked by a long galactic GRB within the past billion years. However, above arguments are from a statistical nature, and actually today there are some candidates for local GRB's Remnants in our galaxy. Additionally, those arguments do not consider the enhancement of star formation and death in the Milky Way, frequently triggered by mergers with smaller galaxies, as the one is seemingly happening now in the other side of our Milky Way. On the other hand, it is not totally clear whether currently there is a star in our galaxy capable of delivering a GRB, but some candidates have been signaled, notably Eta Carinae and WR 104.

There are several hazardous effects that a galactic GRB could cause on our planet: the UV flash from retransmitted gamma radiation in the atmosphere, ozone layer depletion, formation of sunlight absorbing $NO_2$ (with potential global cooling), nitric acid rain, etc. It is challenging to predict the behavior of the biosphere under such perturbation, but above effects seemingly have the potential to cause a mass extinction, especially if the progenitor of the GRB is at one or two kpc from Earth. Actually, some authors have launched the hypothesis of a GRB as initial cause of the Ordovician mass extinction (Melott *et al.*, 2004).

In this work we focus on the short-term effect of the GRB: the UV-flash. The initial gamma radiation extracts electrons from molecules in the atmosphere; these are the so-called primary photoelectrons. The latter, being so energetic, extract other electrons from molecules, which in turn excite molecules and create aurora-like emission. An important fraction of this emission is in the ultraviolet range. Depending mainly on photoabsorption and Rayleigh scattering of the atmospheric molecules, 1–10% of the gamma energy incident at the top of the atmosphere shall reach the ground as this biologically-active "auroral" UV (Smith, Scalo & Wheeler 2004; Martin et al 2009). Also, the irradiances at ground associated to the UV-flash may vary in some percent by the opacity of the other variable atmospheric constituents such as dust, aerosols and clouds. The typical duration of long GRB's can be taken as 10 seconds. Since the lifetimes ($\tau e$) of the upper (excited) molecular electronic states for the most important nitrogen bands are usually shorter ($t_{GRB} \gg \tau e$), then it makes sense to consider that the UV-flash at ground level has the same temporal scale than the GRB. Bear in mind that the interstellar medium is practically transparent to gamma photons, so no delay is to be expected. This is not the case for charged particles, which are deflected by galactic magnetic fields in their travel through space.

One important difference between solar-UV and GRB-retransmitted-UV radiation is that the later delivers a high fluence in a very short time scale. As we show in Fig. 2 below, the smaller the wavelength, the greater the difference between GRB-UV and solar-UV. This difference is amplified when biological action is considered, as smaller wavelengths are much more damaging. For instance, as shown by some of us in Table 3 of (Martin et al 2009), a star as far away as 4,45 kpc, in just 10 seconds can deliver at ground level (with our present atmosphere), the same *effective biological* fluence that the Sun can do in the whole day. Additionally, some fluence of the very deleterious UV-C band will reach the ground in the case of GRB-retransmitted radiation, because part of it is generated close enough to the ground and thus not totally screened.

So far, studies on the action of a GRB on biosphere have focused on surface biota. In this work, we study the potential photosynthesis inhibition of ocean phytoplankton under the influence of the UV-flash coming from a galactic gamma ray burst. Ocean phytoplankton are of enormous importance, at least for two reasons. On one hand, they are involved in 40% (Falkowski 1994) of the photosynthetic activity in the oceans, thus capturing great quantities of carbon dioxide and releasing big amounts of oxygen to the atmosphere. One single species, *Prochlorococcus marinus*, is responsible for 20% of the total oxygen released by the biosphere (Partensky, Hess & Vaulot, 1999). On the other hand, phytoplankton are the starting species of the food assemblages in the ocean, being the basic food for the consumers, thus influencing the biodiversity of the higher levels in the food assemblage. Above facts indicate that a massive perturbation, for any reason, to oceanic phytoplankton, could be transmitted through the food assemblage to higher trophic levels which nutritionally depend on them, and through atmospheric effects to the climate system. This underscores the importance of the investigation of the potential past, present and future effects of a GRB on these groups of species.

## 2 Materials and methods

In this work we have assumed the typical nearest burst in the last billion years as in (Thomas *et al.*, 2005), i. e., at two kpc from Earth and delivering a fluence of 100 kJ/m$^2$ at the top of the atmosphere. To calculate the ultraviolet radiation (UVR) spectrum reaching the ocean's surface, we followed (Martin *et al.* 2009). Then we consider an average ocean albedo of 6,6 %, as in (Cockell 2000). This was used to calculate the UVR spectrum just below the ocean surface $E_0(\lambda, 0^-)$.

In order to analyze the transport of radiation in the ocean, it is necessary to specify the optical quality of the water. This is determined by the absorption and dispersion of photons, and at least four components in ocean water count: phytoplankton, dissolved organic carbon, particulate (non-algal) material and water itself. This information is contained in the spectrum of the attenuation coefficients of radiation *K(λ)* vs. *λ*. According to this, N. Jerlov formally classified oceanic water types (Jerlov 1951). Type I waters were represented by extremely clear oceanic waters. Most clear coastal waters were classified as Type II because attenuation tends to be greater than that for oceanic waters of low productivity. However, many water bodies were found to lie between Types I and II and subsequently were introduced intermediate Types IA and IB (Jerlov 1964). Type III waters are fairly turbid and some regions of coastal upwelling are so turbid that they were not initially classified. Water types I, II, and III, are roughly equivalent to oligo-, meso- and eutrophic waters. For further information, readers are referred to (Jerlov 1976; Shifrin 1988). In this work, we used the values of *K(λ) for* water types I, II, and III; as reported in (Shifrin 1988), i. e., in intervals of 25 nm. Then, linear interpolation was used to obtain the full set of *K(λ)*. Below we show the corresponding plot.

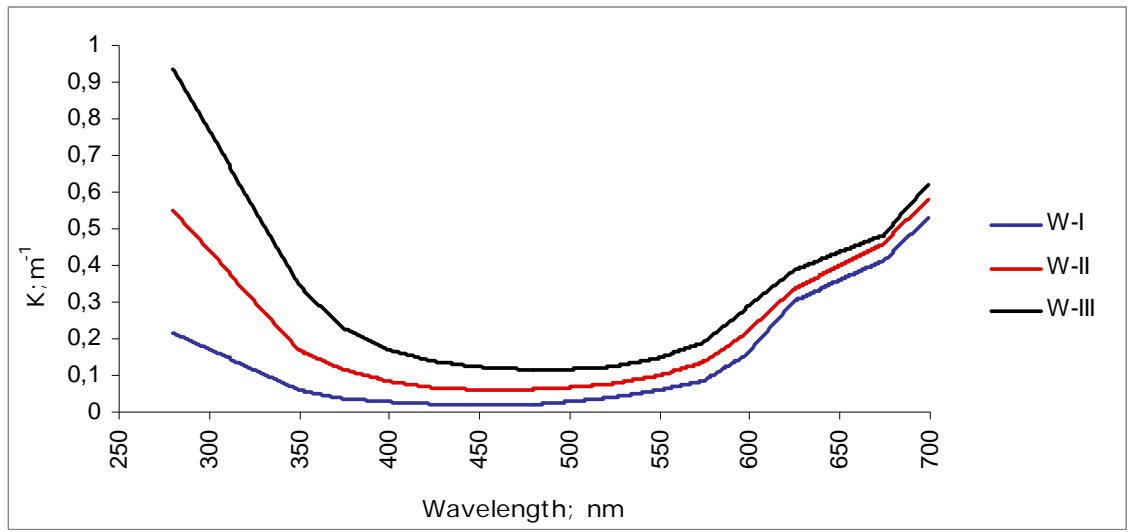

Fig. 1 Attenuation coefficients for Jerlov water types I, II, and III

Not all wavelengths of ionizing electromagnetic radiation have the same influence on cells. Within the UV range, the smaller the wavelength implies the greater the damage, both because of more photon energy and increased absorption probability. Thus, to quantify the biological action of UV, the so-called biological action spectra are obtained for specific damages: DNA damage, inhibition of photosynthesis, etc. These spectra can be plotted as a set of values expressing the biological action *e(λ)* vs. wavelength λ, as done in Fig. 2 in (Cockell 2000). Photosynthesis inhibition by UVR appears as a combination of two factors:
- photoinhibition *per se*: damage in the photosynthetic apparatus, specifically in photosystem II and in the photosynthetic enzyme Rubisco
- DNA damage (because the cell should spend energy repairing DNA, which otherwise would be used in the photosynthesis process)

In this work, we utilized biological action spectra $e(\lambda)$ for phytoplankton inhibition as in (Cockell 2000).

The irradiances and fluences at depth $z$ are given by:

$$E(z) = \sum_{\lambda} E_0(\lambda, 0^-) e^{-K(\lambda)z} d\lambda \quad (1)$$

$$F(z) = E(z)\Delta t \quad (2)$$

where $\Delta t$ is the exposure time to UVR.

The (effective) biological fluxes or dose rates $E^*(z)$ and the (effective) biological fluences or doses $F^*(z)$ at depth $z$ are calculated by:

$$E^*(z) = \sum_{\lambda} e(\lambda) E_0(\lambda, 0^-) e^{-K(\lambda)z} d\lambda \quad (3)$$

$$F^*(z) = E^*(z)\Delta t \quad (4)$$

For the short term effect we used the *H* model for photosynthesis inhibition (Fritz *et al.*, 2008). This model assumes no repair during the brief time interval of the flash (10 seconds), and uses the adimensional effective dose $H^*(z)$ is instead of the dimensional one $F^*(z)$:

$$\frac{P}{P_0} = e^{-H^*(z)} \quad (5)$$

where $P$ and $P_0$ are photosynthesis rates just after and just before the GRB, respectively.

Considering:

$$H^*(z) = H_A^*(z) + H_B^*(z) + H_C^*(z) \quad (6)$$

Substituting (6) in (5):

$$\frac{P}{P_0} = e^{-H_A^*} e^{-H_B^*} e^{-H_C^*} \quad (7)$$

Defining,

$$\frac{P}{P_0} = e^{-H_i^*} \quad (8)$$

where $i = A, B, C$, the effect of each UV band on photosynthesis inhibition is explicit, and equation (8) can be rewritten:

$$\frac{P}{P_0} = \left(\frac{P}{P_0}\right)_A \left(\frac{P}{P_0}\right)_B \left(\frac{P}{P_0}\right)_C \qquad (9)$$

**3 Results and discussion**

**3.1** GRB spectra in the UV wavelength range

We recommend (Smith, Scalo & Wheeler 2004) and references therein to those readers interested in the spectrum of the GRB at the top of the atmosphere. As our main interests in this work start at ocean surface, we present a plot of the GRB spectrum received in the UV wavelength range, and compared it with the daily average solar-UV at mid-latitudes.

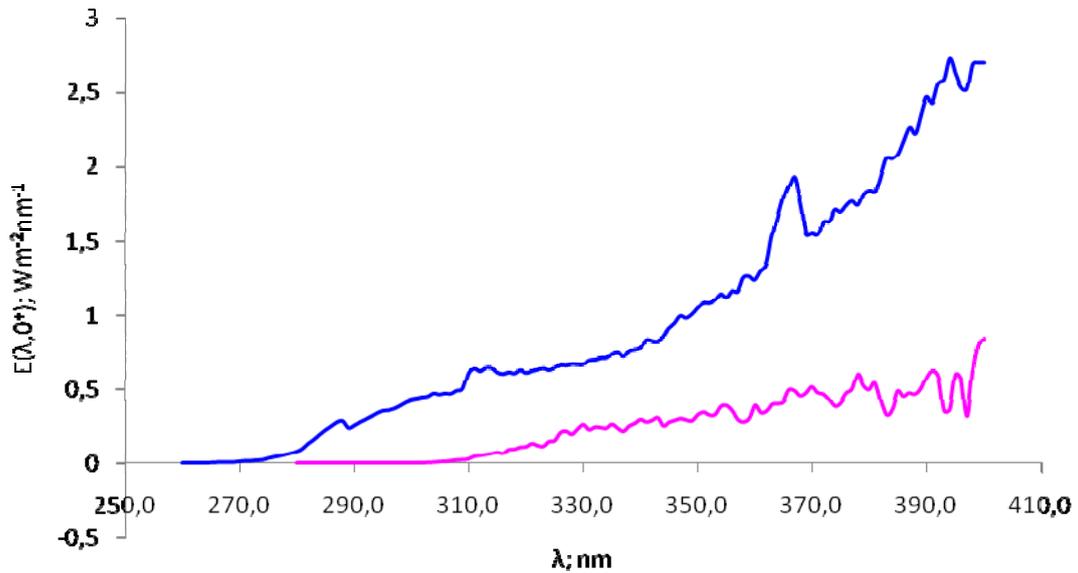

**Fig. 2** GRB spectrum in the UV wavelength range received at ocean surface.

We see that both in the A and B band, the GRB-retransmitted UV have much greater spectral irradiances than the solar UV. There is also some UV-C in the case of the GRB, not very noticeable in the plot due to the scale used, though it can still be seen at 280nm and a bit less.
 The transfer of UV photons in ocean water will depend on water types, so, as an example, we show the UV spectrum at 10 meters depth for the three water types we use in this work. As expected, the attenuation of UV is greater in water type III, then II and then in I.

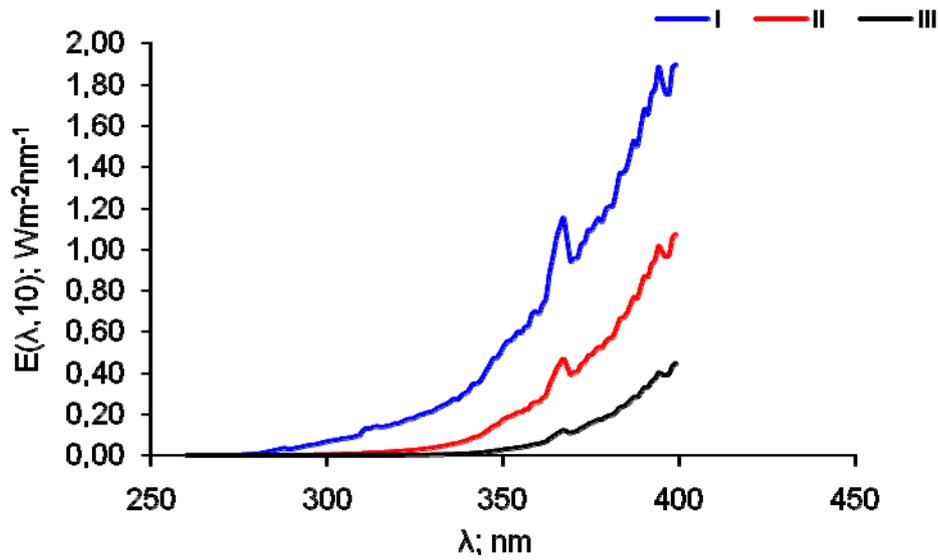

**Fig. 3** GRB spectrum in the UV wavelength range received at 10 meters depth.

3.2 Fluences delivered

During the 10 seconds lasting the UV flash, considerable fluences can be delivered in the water column, depending on the water type, as can be seen in Fig. 4 below. Notice that in the case of the clearest water type I, even at around 100 meters, some fluence can be received.

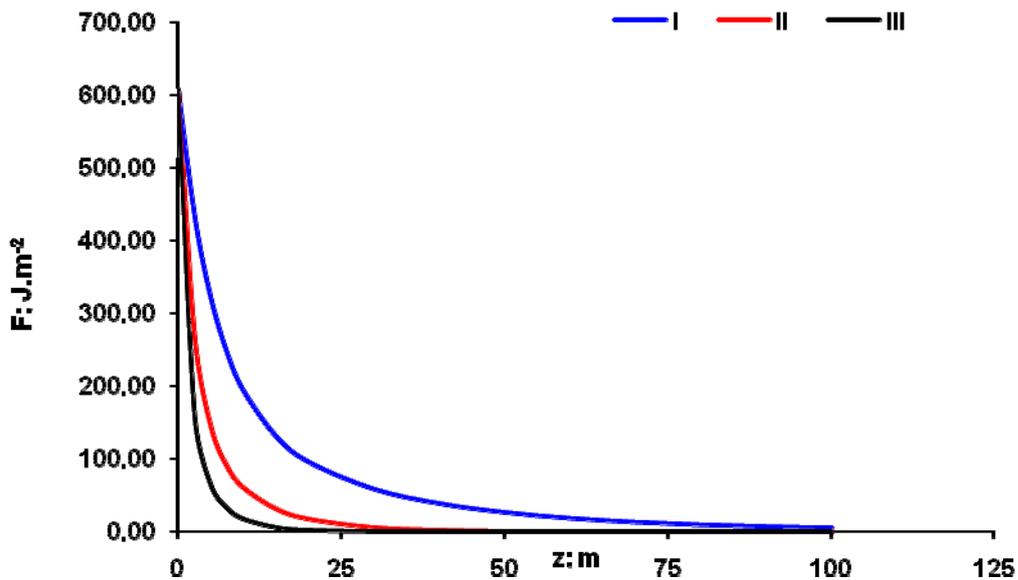

**Fig. 4** Total fluences delivered by the UV flash at the first hundred meters of the water column, for three water types of Jerlov general classification (I, II and III).

As expected, the absorption of UV turns out to be differential; being all three water types more transparent to A photons, followed by the B ones, and ultimately the C band is strongly absorbed, as can be seen in Figs. 5-7.

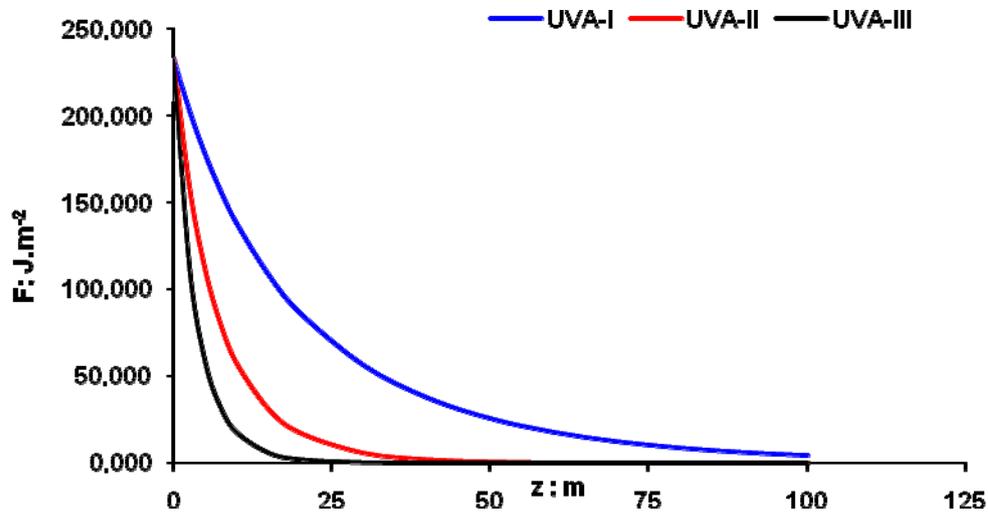

**Fig. 5** Fluence delivered by the UV flash at the water column, for the A band and for three water types of Jerlov general classification (I, II and III).

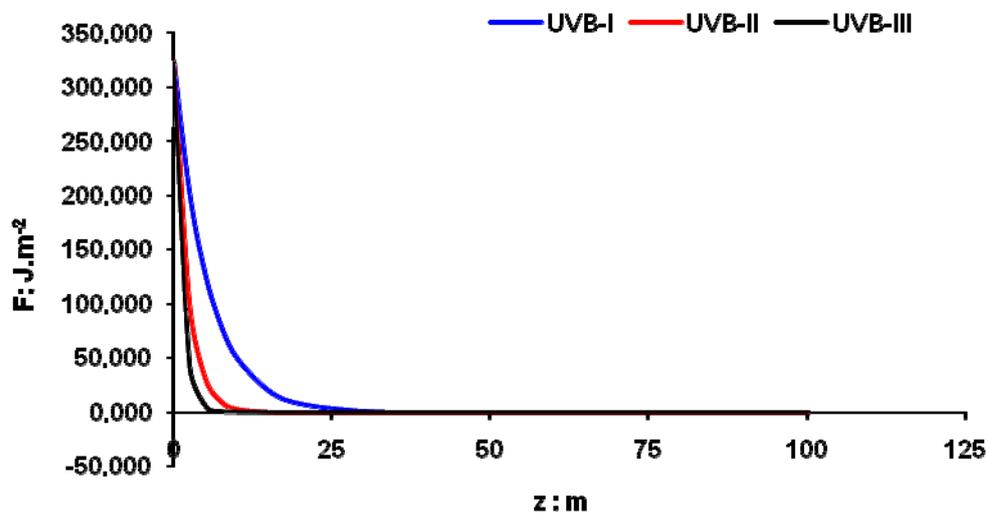

**Fig. 6** Fluence delivered by the UV flash at the water column, for the B band and for three water types of Jerlov general classification (I, II and III).

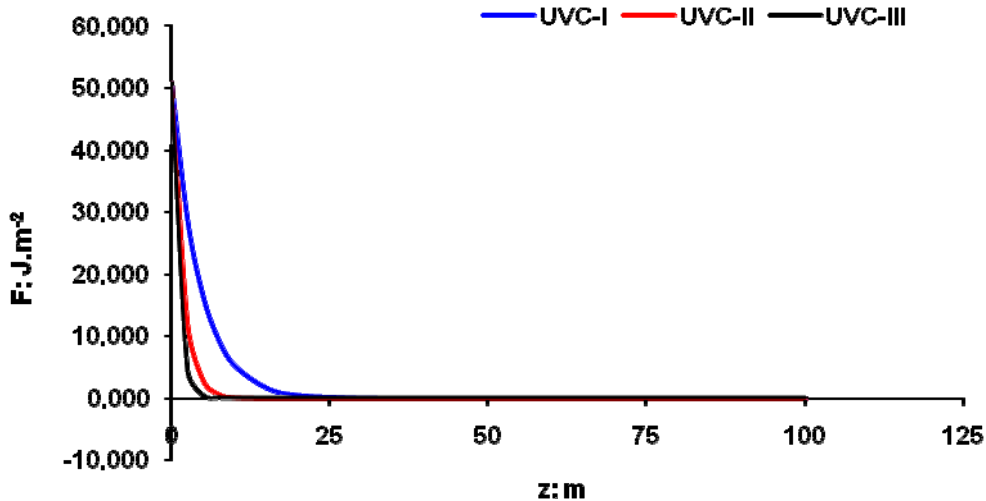

**Fig. 7** Fluence delivered by the UV flash at the water column, for the C band and for three water types of Jerlov general classification (I, II and III).

3.3 The inhibition of photosynthesis

After the ten seconds lasting the UV flash, photosynthesis inhibition of phytoplankton is considerably depleted in the first meters of the water column, as can be seen in Fig. 8 below. It is even totally suppressed down to 80 meters in water type I.

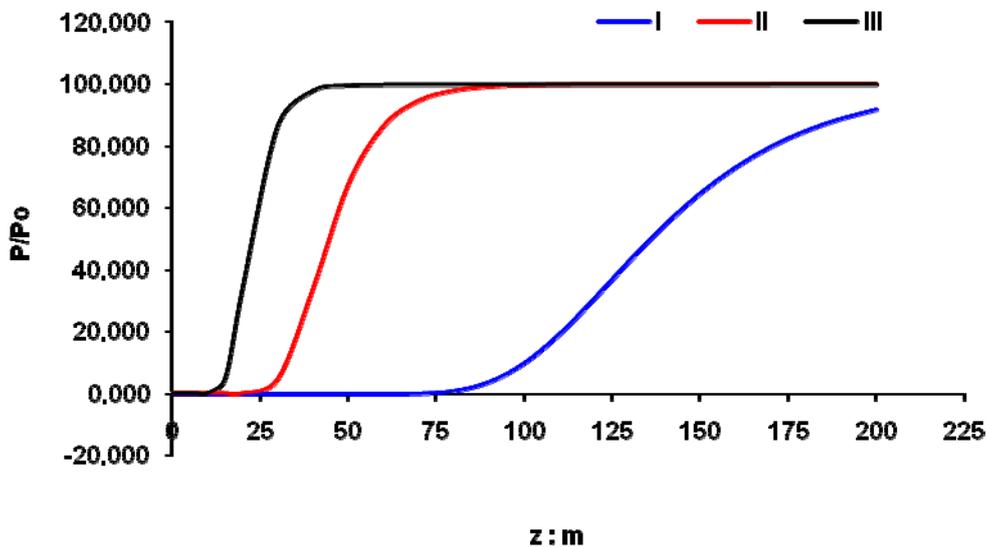

**Fig. 8 Relative p**hotosynthesis rate (%) at the first 200 meters of the water column (photic zone), for three water types of Jerlov general classification (I, II and III). *P* and *$P_0$* are photosynthesis rates just after and just before the GRB, respectively

Figs. 9-11 show that also the inhibition of phytoplankton photosynthesis is band specific, i. e., band A produces the greatest inhibition, followed by B and then by C. There are two competing factors here. A photons are less absorbed by ocean water at any given depth, then B and then C, so more A photons will strike phytoplankton cells,

followed by B and C photons. On the other hand, the photosynthesis inhibition power follows the reverse order (C, B, A). In this case prevails the greatest transparency of water for A photons, so the A band causes the greatest photosynthesis inhibition.

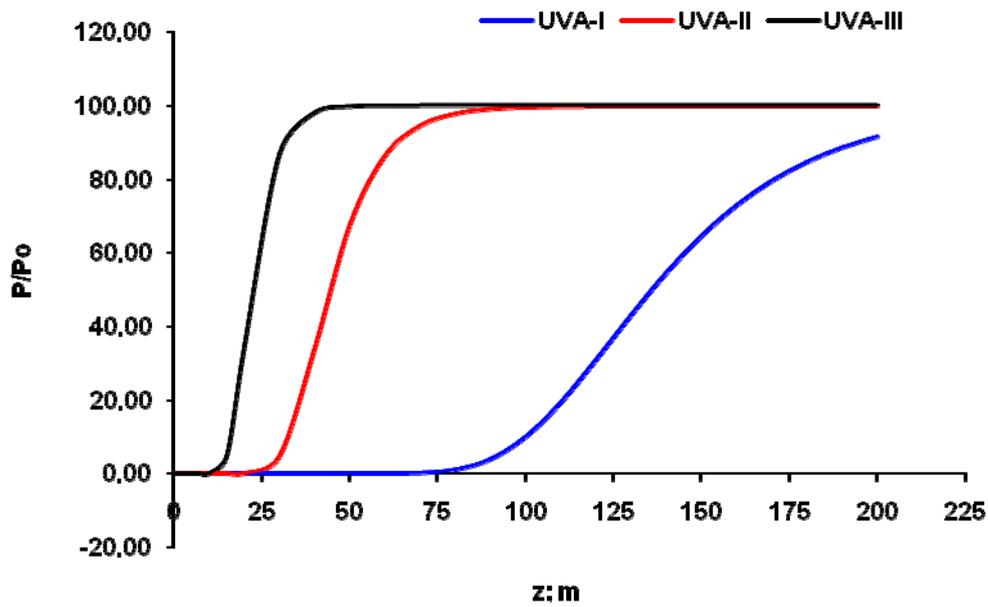

**Fig. 9 Influence of UV-A on relative p**hotosynthesis rate (%) at the first 200 meters of the water column (photic zone), for three water types of Jerlov general classification (I, II and III). *P* and *P₀* are photosynthesis rates just after and just before the GRB, respectively

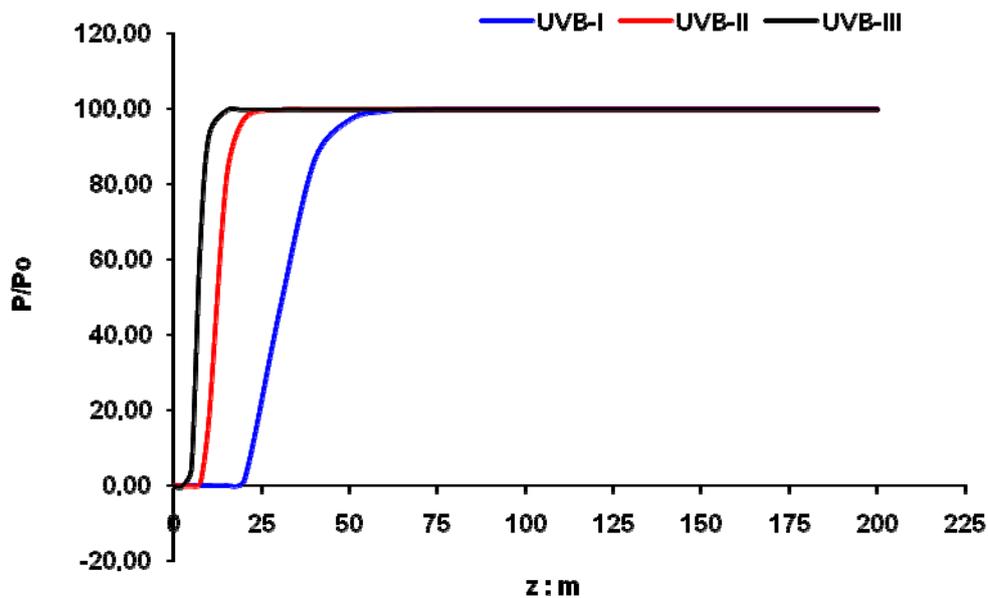

**Fig. 10 Influence of UV-B on relative p**hotosynthesis rate (%) at the first 200 meters of the water column (photic zone), for three water types of Jerlov general classification

(I, II and III). *P* and *P₀* are photosynthesis rates just after and just before the GRB, respectively

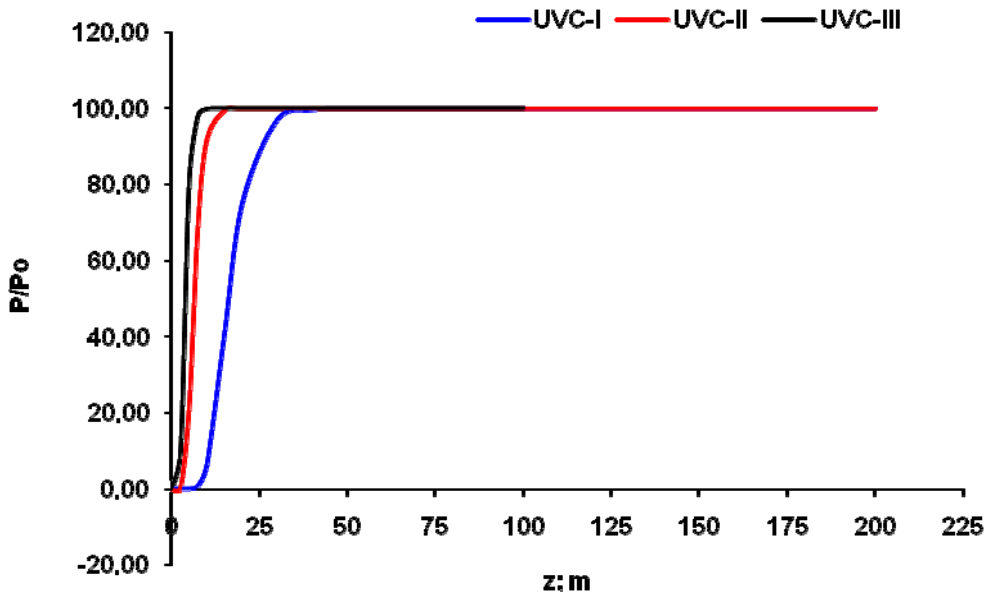

**Fig. 11 Influence of UV-C on relative p**hotosynthesis rate (%) at the first 200 meters of the water column (photic zone), for three water types of Jerlov general classification (I, II and III). *P* and *P₀* are photosynthesis rates just after and just before the GRB, respectively

4 Conclusions

As show our calculations and plots above, a gamma ray burst originated as far as two kpc (more than 6000 light years) from us, can deliver appreciable dosages of ultraviolet radiation in a considerable part of the photic zone. The clearest the water, the greatest the dosage, as expected. A night flash could probably cause more damage in phytoplankton because cell division is synchronised in natural oceanic phytoplankton occurring during night hours (e.g. Agawin and Agustí 2005). Additionally, it would affect several organisms used to be in deep waters during daylight time, not adapted to crude photobiological regimes.

Above UV irradiances and fluences are amplified when biological damage is considered, thus the so called *effective biological fluences* can really suppress photosynthesis in the first 20 or 30 meters of the water column for turbid or intermediate waters, and even down to 75 meters for clear waters. However, we should bear in mind that in this photosynthesis model we do not consider the existing molecular mechanisms for repairing the UV damage, as the flash only lasts ten seconds. Once the flash finishes, it makes sense to consider repair mechanisms, which will determine the time for photosynthesis to recover to usual levels. Indeed, the whole issue of recovery after a nearby galactic GRB strikes is complicated, because long-term effects such as the depletion of the ozone layer (allowing reaching the ground more solar UV) will become important after the brief UV-flash. That is, the photosynthesis recovery would have to be done under the influence of more solar-UV during a decade or so. This shall be presented in a forthcoming publication.

Other important effect of the UV-flash from a galactic GRB would be DNA damage and cell mortality of the most sensitive oceanic species as *Prochlorococcus* sp. (Llabrés and Agustí 2006, Agustí and Llabrés 2007). Actually, due to the presence of UV photons in the very deleterious C band, we guess that there will be significant phytoplankton mortality due to DNA damage, and the formation of toxic reactive oxygen species in cells. Cell mortality of other planktonic groups like heterotrophic bacteria, and protozoa, that appear also to be highly UV sensitive (Ferreyra *et al*., 2006) may also occur. This could be transmitted through the trophic assemblage and therefore affect many other groups.

**References**

Agawin, N.S.R., Agustí, S.: *Prochlorococcus* and Synechococcus cells in the Central Atlantic ocean: distribution, growth and mortality grazing rates. Vie et Milieu **55**: 165-175 (2005)

Agustí, S., Llabrés, M.: Solar Radiation-induced Mortality of Marine Pico-phytoplankton in the Oligotrophic Ocean. Photochemistry and Photobiology **83**: 793–801 (2007)

ASTM G173 - 03e1. ASTM G173 - 03e1 Standard Tables for Reference Solar Spectral Irradiances. http://www.astm.org/Standards/G173.htm

Cockell, C.: Ultraviolet radiation and the photobiology of Earth's early oceans. Orig. Life Evol. Biospheres **30**, 467–499 (2000)

Ferreyra G.A., Mostajir, B., Schloss, I.R., Chatila, K., Ferrario, M.E., Sargian, P., Roy, S., Prod'homme, J., Demers, S.,: Ultraviolet-B radiation effects on the structure and function of lower trophic levels of the marine planktonic food web. Photochem. Photobiol. **82**, 887-897 (2006)

Falkowski, P.G.: The role of phytoplankton photosynthesis in global biogeochemical cycles. Photosynthesis Research **39**: 235-258 (1994)

Fritz, J., Neale, P., Davis, R., Peloquin, J.: Response of Antarctic phytoplankton to solar UVR exposure: inhibition and recovery of photosynthesis in coastal and pelagic assemblages. Mar Ecol Prog Ser, 365, 1-16 (2008). doi: 10.3354/meps07610

Jerlov, N. G.: Optical Studies of Ocean Water. Report of Swedish Deep-Sea Expeditions 3, 73–97 (1951).

Jerlov, N. G.: Optical Classification of Ocean Water. In Physical Aspects of Light in the Sea. (Honolulu:University of Hawaii Press), pp. 45–49 (1964).

Jerlov, N. G.: Applied Optics. (Amsterdam: Elsevier Scientific Publishing Company) (1976).

Llabrés, M., Agustí, S.: Picophytoplankton cell death induced by UV radiation: Evidence for oceanic Atlantic communities Limnol. Oceanogr. **51**(1), 21–29 (2006)

Martin, O., Galante, D., Cardenas, R., Horvath, J.: Short-term effects of gamma ray bursts on Earth. Astrophys Space Sci **321**: 161–167 (2009). DOI 10.1007/s10509-009-0037-3

*Melott, A., B. Lieberman, C. Laird, L. Martin, M. Medvedev, B. Thomas, J. Cannizzo, N. Gehrels, C. Jackman:* Did a gamma-ray burst initiate the late Ordovician mass extinction? Int.J.Astrobiol.**3**:55 (2004)

Partensky, F., Hess, W., Vaulot, D.: *Prochlorococcus*, a Marine Photosynthetic Prokaryote of Global Significance, Microbiology and Molecular Biology Reviews, **63**, 106 (1999)

Smith, D. S., Scalo, J., Wheeler, J.C.: Transport of Ionizing Radiation in Terrestrial-like Exoplanet Atmospheres. Icarus **171**, 229-253 (2004)

Shifrin, K., 1988, ''Physical Optics of Ocean Water'', American Institute of Physics, New York

Thomas, B., Melott, A., Jackman, C., Laird, C., Medvedev, M., Stolarski, R., Gehrels, N., Cannizzo, J., Hogan, D., Ejzak, L.: Gamma-Ray Bursts and the Earth: Exploration of Atmospheric, Biological, Climatic and Biogeochemical Effects. Astrophys.J. **634**, 509-533 (2005)